\documentclass{article}

\newcommand{\Id}{\mathbf{1}}
\newcommand{\e}{\mathfrak{e}}

\newcommand{\Eq}[1]{Eq.~(\ref{#1})}
\newcommand{\Sec}[1]{Sec.~\ref{Sec:#1}}
\newcommand{\Cl}{C\!\ell}
\newcommand{\C}{\mathbb{C}}
\newcommand{\R}{\mathbb{R}}
\newcommand{\Q}{\mathbb{H}}
\newcommand{\Z}{\mathbb{Z}}
\newcommand{\qu}[1]{\mathfrak{#1}}
\newcommand{\qi}{\qu{i}}
\newcommand{\qj}{\qu{j}}
\newcommand{\qk}{\qu{k}}
\newcommand{\bi}{\boldsymbol\iota}
\newcommand{\di}{\breve\bi}
\newcommand{\hotimes}{\mathbin{\widehat{\otimes}}}
\newcommand{\isom}{\simeq}

\newcommand{\es}{\mathfrak{s}}

\newcommand{\mat}[1]{\begin{pmatrix}#1\end{pmatrix}}
\newcommand{\bmat}[1]{\begin{bmatrix}#1\end{bmatrix}}

\newcommand{\ket}[1]{| #1 \rangle}
\newcommand{\bra}[1]{\langle #1 |}

\usepackage{amsmath,amssymb,bm,amsthm}
\DeclareMathOperator{\Tr}{Tr} 
\usepackage{verbatim}

\theoremstyle{definition}
\newtheorem{definition}{Definition} 
\begin{document}

\tolerance=5000
\binoppenalty=2500
\relpenalty=2000

\normalsize 
\thispagestyle{empty}
\setcounter{page}{1}

\vspace*{0.88truein}

\centerline{\bf Quantum Circuits and $\bm{Spin(3n)}$ Groups} 
\vspace*{0.37truein}
\centerline{\footnotesize
Alexander Yu.\ Vlasov}
\vspace*{0.015truein}
\centerline{\footnotesize\it Federal Radiology Center (IRH)}
\centerline{\footnotesize\it 197101,Mira Street 8, St.--Petersburg, Russia}
\centerline{\footnotesize\it A. Friedmann Laboratory for Theoretical Physics}
\centerline{\footnotesize\it 191023, Griboedov Canal 30/32, St.--Petersburg, Russia}

\vspace*{0.21truein}

\begin{abstract}
 All quantum gates with one and two qubits may be described
 by elements of $Spin$ groups due to isomorphisms 
 $Spin(3) \isom SU(2)$ and  $Spin(6) \isom SU(4)$.  
 However, the group of $n$-qubit gates $SU(2^n)$ for $n > 2$ 
 has bigger dimension than $Spin(3n)$. 
 A quantum circuit with one- and two-qubit gates may be 
 used for construction of arbitrary unitary transformation $SU(2^n)$. 
 Analogously, the `$Spin(3n)$ circuits' are introduced in this work as
 products of elements associated with one- and 
 two-qubit gates with respect to the above-mentioned isomorphisms.
 
 The matrix tensor product implementation of the $Spin(3n)$ group
 together with relevant models by usual quantum circuits with 
 $2n$ qubits are investigated in such a framework.
 A certain resemblance with well-known sets of non-universal 
 quantum gates ({\em e.g.}, matchgates, noninteracting-fermion 
 quantum circuits) related with $Spin(2n)$ may be found in 
 presented approach. Finally, a possibility of the classical 
 simulation of such circuits in polynomial time is discussed.
\end{abstract}

\vspace*{10pt}

{{\em Keywords:} quantum computation, matchgates, spin groups,polynomial time}
\vspace*{3pt}

\vspace*{1pt} 


\begin{flushright}
03.67.Ac, 02.20.Sv, 03.65.Fd, 89.70.Eg
\end{flushright}





\section{Introduction}
\label{Sec:intro}

An example of representation of quantum gates with $Spin$ groups 
and Clifford algebras was considered in earlier work
\cite{Vla0}. Similar approach was also discussed 
due to relation of {\em matchgates} and 
{\em noninteracting-fermion} 
quantum circuits \cite{Val1,TD2,Kni1,Joz8,Joz9,JM13}.
It was also found in a broader context, that the $Spin$ groups 
can be also related with so-called {\em holographic algorithms}
\cite{Val7,Lan13}, but this issue is beyond the 
scope of the presented work. 

Such a representation corresponds to the non-universal set
of quantum one- and two-qubit gates generating the
subgroup isomorphic to $Spin(2n)$ for a quantum circuit with
$n$ qubits.
It was shown directly in \cite{Vla0} 
and also follows from other works \cite{TD2,Joz8,Joz9} 
due to definition of $Spin(2n)$ \cite{ClDir,Port,Post}.

Relation with the physical fermions is not obvious from
such a construction with $Spin(2n)$, {\em e.g.}, for 
one qubit $Spin(2)$ is simply one-parameter group. 
On the other hand, there
is an isomorphism between $Spin(3)$ and the group of
one-qubit gates $SU(2)$ and it has the direct relation 
with a physical implementation of a single qubit by a 
spin-half particle.

The isomorphism \cite{Port,Post} between $Spin(6)$ and 
the group of two-qubit gates $SU(4)$ is less trivial and does
not have clear physical implications. A similar relation
between group of $n$-qubit gates $SU(2^n)$ and $Spin(3n)$ 
{\em may not exist} for $n>2$, because dimensions of such 
groups are $4^n-1$ and $3n(3n-1)/2$ respectively.

\smallskip

Let us consider $3n$ elements $\e_j$, $j=1,\ldots,3n$
(used further for construction of the Clifford 
algebra $\Cl(3n)$ \Eq{Cln} and the $Spin(3n)$ group)
together with the subdivision:
\begin{equation}
\e^{(l)}_\nu = \e_{3(l-1)+\nu},
\label{esub}
\end{equation}
where $l=1,\ldots,n$ and $\nu=1,2,3$. 

An idea to identify $(l)$ with the index of a line 
in some circuit might be more clear 
further in \Sec{StrCl3n} with rewriting 
$\e^{(l)}_\nu$ as \Eq{ehot} and in \Sec{QCirc}
due to possibility to associate each $(l)$
with pair of lines in quantum circuits 
with $2n$ qubits.

A set with $k$ indexes
$l_1<l_2< \cdots < l_k$ corresponds to $3k$ elements
$\e^{(l_1)}_\nu,\ldots,\e^{(l_k)}_\mu$ those may be used
for construction of subalgebra $\Cl(3k) \subset \Cl(3n)$
and subgroup $Spin(3k) \subset Spin(3n)$.
An element of the subgroup is considered further as an analogue 
of a gate with $k$ lines. 
For $k=1,2$ there are isomorphisms with the groups of quantum
one- and two-qubit gates, because $Spin(3) \isom SU(2)$ and
$Spin(6) \isom SU(4)$.

The composition of such analogues of one- and two-line gates 
may be considered as some `$Spin(3n)$ circuits.'
Formally, any quantum circuit composed only from one-
and two-qubit gates would define a `$Spin(3n)$ circuits' 
by remapping of all such gates using above-mentioned 
isomorphisms into elements of the group $Spin(3n)$.

Despite of the formal isomorphisms for gates 
with fixed one or two lines, results of composition 
with different lines may not be isomorphic, because quantum 
one- and two-qubit gates may produce whole group $SU(2^n)$,
unlike analogous gates in `$Spin(3n)$ circuits.'


\paragraph*{The plan of the paper.}
In the \Sec{SPIN3N} construction of 
`Spin(3n) circuits' is discussed using rather abstract
and general mathematical structures.
Results and methods from the \Sec{SPIN3N} are 
revisited and become more descriptive in the next two sections. 
The \Sec{matr} uses more understanding 
models with matrices and \Sec{QCirc} 
illustrates some important results 
using quantum circuits with $2n$ qubits.

Finally, in \Sec{ClSim} the effective simulation of 
`$Spin(3n)$ circuits' by classical computers is 
discussed using the model with $2n$ qubits from \Sec{QCirc}
and methods of simulating quantum circuits with matchgates 
developed earlier \cite{Joz8,Joz9}.

\section{Structure of Spin(3n) groups}
\label{Sec:SPIN3N}
\subsection{Clifford algebras and Spin groups}
\label{Sec:ClSpin}

Let us recall some preliminaries
\cite{ClDir,Port,Post,KM,VDW}. Real Clifford algebra
is defined by $n$ generators $\e_j$ with properties
\begin{equation}
 \e_j^2 = -\Id, \qquad \e_j\e_k = -\e_k\e_j \quad (j \ne k),
\label{Cln}
\end{equation}
often written in a single equation 
$\e_j\e_k + \e_k\e_j = -2\delta_{jk}\Id$.
Different products of the generators $\e_j$ is a basis of
the {\em (universal) real Clifford algebra} $\Cl(n)$ 
with dimension $2^n$. 

A similar equation $\e_j\e_k+ \e_k\e_j = 2\delta_{jk}\Id$
defines $\Cl_+(n)$.
In the most general case a real Clifford algebra is defined
by a quadratic form $g$ with the matrix $g_{jk}$
\begin{equation}
 \e_j\e_k + \e_k\e_j = -2g_{jk}\Id.
\label{Clg} 
\end{equation}
For a diagonal matrix $g$ with the signature $(m,n-m)$
\Eq{Clg} defines the Clifford algebra $\Cl(m,n-m)$, 
{\em e.g.}, $\Cl(n) = \Cl(0,n)$.

The {\em Clifford conjugation} may be
defined on the generators by the equation
\begin{subequations}
\label{clcnj}
\begin{equation}
 \bar{\e}_j\e_j = \Id
 \label{cnjek}
\end{equation}
and extended on the whole algebra by the property
\begin{equation}
 \overline{ab} = \bar{b}\bar{a}.
 \label{cnjab}
\end{equation}
\end{subequations}

The complex Clifford algebra $\Cl(n,\C)$ also is defined 
by \Eq{Cln}, but in such a case 
any signatures are equivalent due to 
the possibility of substitutions $\e_k \mapsto i \e_k$.

Let us consider the $n$-dimensional subspace $V$ of the Clifford
algebra $\Cl(n)$ with elements
\begin{equation}
 \bm{v} \in V,\quad  
 \bm{v} = \sum_{k=1}^n v_k \e_k.
 \label{vsum}
\end{equation}
By {\em the definition} \cite{ClDir,Port,Post} the $Spin(n)$ group 
is generated by all possible products with {\em even} 
number of $\bm{v} \in V$ normalized by the condition 
\begin{equation}
 \sum_{k=1}^n v_k^2 = 1.
 \label{vsqr}
\end{equation}

Real Clifford algebras $\Cl_+(n)$ also
may be used for an analogous definition of the $Spin(n)$ group,
yet $\Cl_+(n)$ and $\Cl(n)$ are different
algebras \cite{ClDir,Port}. The property is important for
some constructions below.

{\em A basic property of the Spin groups}: 
Let $\bm{S} \in Spin(n)$, $\bm{v} \in V$ \Eq{vsum} and
\begin{equation}
\bm{v}' = \bm{S} \bm{v} \bm{S}^{-1},
\label{SvS}
\end{equation} 
then $\bm{v}' \in V$ and
\begin{equation}
 \bm{v}'= \sum_{k=1}^n v'_k \e_k, \quad
 v'_k = \sum_{j=1}^n R_S^{kj} v_j, \quad
 R_S \in SO(n).
 \label{SpinSO}
\end{equation}
Any rotation $R \in SO(n)$ may be represented
in such a way. Due to \Eq{SvS} both $\bm{S}$
and $-\bm{S}$ correspond to the same $R_S \in SO(n)$
and so, it is $2{\rightarrow}1$ {\em covering homomorphism}. 

Thus, dimensions of $Spin(n)$ and $SO(n)$
are the same, $n(n-1)/2$.
On the other hand, any product with even number 
of $\e_k$ belongs to $Spin(n)$. The linear span 
of such products has dimension $2^{n-1}$ and
corresponds to an {\em even subalgebra} of $\Cl(n)$
denoted further as $\Cl^0(n)$.

Due to such a complicated structure of the $Spin$ groups it may be 
more convenient sometimes to use Lie algebras $spin(n)$. 
The elements of $spin(n)$ are linear combinations of 
products $\e_j\e_k$ equipped with the Lie bracket operation 
\begin{equation}
 [a,b] = ab - ba.
 \label{comm}
\end{equation}
The $spin(n)$ is isomorphic with the Lie algebra 
$so(n)$ of the orthogonal group $SO(n)$ \cite{ClDir,Post}.

\subsection{Spin(3) group}
\label{Sec:Spin3}

Clifford algebras with three generators may be 
considered as building blocks in many constructions 
used in this work. The $Spin(3)$ group may be constructed 
both from $\Cl(3)$ and $\Cl_+(3)$, but the algebras 
are not equivalent. 
In both cases an element of the $Spin$ group is represented as
\begin{equation}
 r_0\Id + r_1\e_{23} + r_2\e_{31} + r_3\e_{12},
 \quad r_k \in \R,\quad\sum_{k=0}^3 r^2_k = 1,
 \label{hejk}
\end{equation}
where $\e_{jk} \equiv \e_j\e_k$.
Both for $\Cl(3)$ and $\Cl_+(3)$
\begin{equation}
 (\e_{jk})^2 = \e_j\e_k\e_j\e_k = -\e_j\e_j\e_k\e_k =
 -\e^2_j\e^2_k = -\Id.
 \label{ejkjk}
\end{equation}

Let us recall that {\em the quaternions} $\Q$
\cite{ClDir,Port,Post,KM,VDW} are defined by relations
\begin{multline}
 \qu{q} = q_0\Id + q_1 \qi + q_2 \qj + q_3 \qk,
 \quad q_k \in \R, \\
 \qi\qj = -\qj\qi = \qk,~
 \qk\qi = -\qi\qk = \qj,~
 \qj\qk = -\qk\qj = \qi,\quad
 \qi^2 =\qj^2=\qk^2 = -\Id.
\label{qu}
\end{multline}

\begin{list}{}{}
\item[{\em Note}\/:]
The quaternions formally correspond to {\em an universal} Clifford
algebra $\Cl(2)$ with only {\em two} generators $\qi$, $\qj$ 
and $\qk = \qi\qj$. Omitting the claim about universality
they are treated sometimes as a Clifford algebra with
{\em three} quaternionic units $\qi$, $\qj$, $\qk$
satisfying \Eq{Cln}.
\end{list}

The multiplicative norm of a quaternion $|\qu{q}|$ is defined as
\begin{equation}
 |\qu{q}|^2 = q_0^2 + q_1^2 + q_2^2 + q_3^2.
\label{qn}
\end{equation}
The group of quaternions with the unit norm is isomorphic
with $SU(2)$ \cite{ClDir,Port,Post,KM,VDW}.

The substitutions
\begin{equation}
\qi=\e_{23},\quad
\qj=\e_{31},\quad
\qk=\e_{12}
\label{quejk}
\end{equation}
in \Eq{hejk} are in agreement with properties \Eq{qu} of quaternions. 
They relate to the quaternionic representation 
of group $SU(2) \isom Spin(3)$ and to the isomorphisms
$\Cl^0(3) \isom \Cl_+^0(3) \isom \Q \isom \Cl(2)$.

\smallskip

A representation of the algebra $\Cl(3)$ may be described
by {\em the double quaternions} $^2\Q$ 
with generators expressed as the matrices 
\begin{equation}
 \e_1 = \mat{\qi&0\\0&-\qi},\quad
 \e_2 = \mat{\qj&0\\0&-\qj},\quad
 \e_3 = \mat{\qk&0\\0&-\qk}.
\label{Cl3} 
\end{equation}

\smallskip

The definition below is useful  
for description of $\Cl_+(3)$. 

\begin{definition} 
 The decomplexification \cite{KM} 
 (realification) $\mathcal{A}_\Re$ of the $n$-dimensional 
 complex algebra $\mathcal{A}$ with the basis 
 $a_k$, $k=1,\ldots,n$ is the $2n$-dimensional 
 real algebra with the basis  
 $b_k = a_k$, $b_{k+n} = i a_k$. 
\end{definition} 

The eight-dimensional real Clifford algebra $\Cl_+(3)$
is isomorphic with the {\em decomplexification} of
the Pauli algebra $M(2,\C)$ of complex $2\times 2$ matrices,
 $\Cl_+(3) \isom M(2,\C)_\Re$.
It may be considered in such a way 
due to the commutative element
\begin{equation}
 \bi = \e_{123} \equiv \e_1\e_2\e_3, \quad
 \bi^2 = -\Id, \quad \bi\e_k = \e_k\bi
\label{bi} 
\end{equation}
corresponding to an imaginary unit.
A similar method does not work with $\Cl(3)$ where 
$(\e_1\e_2\e_3)^2 = 1$. 

\smallskip

It may be written
\begin{equation}
 \forall \bm{z} \in \Cl_+(3), \quad
 \bm{z} = \bm{x} + \bi\bm{y}, \quad 
 \bm{x},\bm{y} \in \Cl_+^0(3) \isom \Q.
\label{zxbiy} 
\end{equation}

\Eq{zxbiy} illustrates isomorphisms
\begin{equation}
 M(2,\C)_\Re \isom \Cl_+(3) \isom \C_\Re \otimes \Q.
 \label{CxH}
\end{equation}

In such a complex representation the Pauli matrices
\begin{equation}
 \sigma_1 = \mat{0&1\\1&0},\quad
 \sigma_2 = \mat{0&-i\\i&0},\quad
 \sigma_3 = \mat{1&0\\0&-1}
\label{PauliMat}
\end{equation} 
may be used as generators of the algebra $\Cl_+(3)$
\begin{equation}
 \e_1 = \sigma_1,\quad
 \e_2 = \sigma_2,\quad
 \e_3 = \sigma_3.
\label{Cl3p} 
\end{equation}

Finally, elements of  $\Cl_+^0(3) \isom \Q$
used for construction of $Spin(3)$ \Eq{hejk} may be
written as
\begin{equation}
 \es_1 \equiv \e_{23} = i\sigma_1,~
 \es_2 \equiv \e_{31} = i\sigma_2,~
 \es_3 \equiv \e_{12} = i\sigma_3.
\label{es} 
\end{equation}
The elements \Eq{es} are basis of the Lie algebra
$spin(3)$ and correspond to the representation
of $spin(3) \isom su(2)$ via {\em anti-Hermitian}
matrices. It may be checked that {\em the Hermitian conjugate}
of the matrix in such a representation is in agreement
with the Clifford conjugation \Eq{clcnj}, 
$\bar{\e}_{jk} = -{\e}_{jk}$.

\begin{list}{}{}
\item[{\em Note}\/:]
For construction of the group $SU(n)$ in physical applications 
{\em Hermitian} matrices $H^\dag=H$ are often used together 
with the equation $U = e^{iH}$ for $U \in SU(n)$ group
and the commutator $i[A,B]$ instead of \Eq{comm}. 

Sometimes it may produce some difficulties, 
{\em e.g.}, in construction of $SU(2)$ with
quaternions defined by relations \Eq{qu} without an
element representing a commutative 
imaginary unit.

So, the more common description with the relation between 
Lie algebras and Lie groups expressed as $G = e^{A}$
may be appropriate.
In such a definition the Lie algebra $su(n)$ is represented
by {\em anti-Hermitian} matrices $A^\dag = -A$ 
with Lie brackets \Eq{comm} and the relation
$U=e^{A}$ for $U \in SU(n)$.
\end{list}

\subsection{Different tensor products}
\label{Sec:Tens}

The consideration of $\Cl_+(3)$ as the Pauli algebra of $2 \times 2$ 
complex matrices may be used for the description of
quantum gates \cite{HavDor2}. 

The group of quantum $n$-qubit gates $SU(2^n)$
may be expressed using $2^n \times 2^n$ complex
matrices $M(2^n,\C)$ represented in turn as 
{\em the complex tensor product}
\begin{equation}
 M(2^n,\C) = \underbrace{M(2,\C) \otimes \cdots \otimes M(2,\C)}_n.
 \label{oxM2C}
\end{equation}

Due to \Eq{CxH} 
an analogue of \Eq{oxM2C} for {\em the real tensor product}
may be written
\begin{equation}
 M(2^n,\C)_\Re \isom 
 \C_\Re \otimes \underbrace{\Q \otimes \cdots \otimes \Q}_n
 = \C_\Re \otimes \Q^{\otimes n}.
\label{MCHn}
\end{equation}

Some subtleties may exist here, because a tensor product
of 8D real algebras $\Cl_+(3)$ has the real dimension  
$8^n{=}2^{3n}$, but $M(2^n,\C)$ has dimension
$2^{2n}$ as the complex algebra and $2^{2n+1}$
as the real one. 

Let us recollect a method to get rid of the extra
dimensions \cite{HavDor2}. 
Each term in the real tensor product of $\Cl_+(3)$ has its
own imaginary unit
\begin{equation}
 \bi_k = {\underbrace{\Id\otimes\cdots\otimes \Id}_{k-1}\,}\otimes
  \bi\otimes\underbrace{\Id\otimes\cdots\otimes\Id}_{n-k},
\label{bik}  
\end{equation}
where $\bi$ is defined by \Eq{bi}.
It is possible to introduce a projector
\begin{equation}
 P_\iota = \frac{1}{2^{n-1}}\prod_{k=2}^n (1 - \bi_1\bi_k).
 \label{Pc}
\end{equation}
It is called {\em the correlator} in \cite{HavDor2} with the
properties 
\begin{equation}
\bi_k P_\iota = \bi_1 P_\iota
\Longrightarrow 
\bi_k \bi_j P_\iota = - P_\iota, 
\label{iPc}
\end{equation} 
so, all the complex units may be `aggregated' by the projector.

In fact, the method is very general and may be used
for tensor products of real spaces with complex structures
of any dimension.

\smallskip

However, it may not be applied
to the description of the structure of $Spin(3n)$ groups and 
Clifford algebras $\Cl_+(3n)$ discussed in this section. 
Indeed, the method works because all $\bi_k$ are commutative 
$\bi_k\bi_j = \bi_j\bi_k$ due to the definition \Eq{bik}, but
for the description of $Spin(3n)$ circuits instead of $\bi_k$
should be used
\begin{equation}
\bi^{(k)} = \e_{123}^{(k)} \equiv \e_1^{(k)}\e_2^{(k)}\e_3^{(k)},
\label{bip}
\end{equation}
where $\e_\nu^{(k)}$ are defined by \Eq{esub}. 
Unlike $\bi_k$, elements $\bi^{(k)}$ anticommute:
$\bi^{(k)}\bi^{(j)}=-\bi^{(j)}\bi^{(k)}$, $ k \ne j$ and
a product such as \Eq{Pc} is {\em not} a projector
with the desired properties \Eq{iPc}.

Despite of this problem, the Clifford algebra $\Cl_+(3n)$
as a real linear space may be constructed
as a tensor product of $n$ copies of $\Cl_+(3)$.
It follows from the general property of Clifford
algebras \cite{ClDir,Post}.
The difference with the usual tensor product 
is the definition of multiplication discussed further.

\begin{definition}
An algebra $\mathcal{A}$ is called
{\em $\Z_2$-graded}, if it may be decomposed into
the direct sum of even and odd linear subspaces
$\mathcal{A} = \mathcal{A}^0 \oplus \mathcal{A}^1$
with the property 
\begin{equation}
a \in \mathcal{A}^j,~ 
b \in \mathcal{A}^k \Longrightarrow
ab \in \mathcal{A}^{(j + k) \bmod 2}. 
\label{gradab}
\end{equation}
Here only an even subspace $\mathcal{A}^0$ is a subalgebra.
\end{definition}

\begin{definition}
For two $\Z_2$-graded algebras 
$\mathcal{A} = \mathcal{A}^0 \oplus \mathcal{A}^1$
and $\mathcal{B} = \mathcal{B}^0 \oplus \mathcal{B}^1$
{\em the $\Z_2$-graded tensor product}
is defined 
\begin{equation}
 (a \otimes b) (a' \otimes b') =
 (-1)^{jk} (aa') \otimes (bb')
\label{gtens} 
\end{equation}
if $b \in \mathcal{B}^j$ and $a' \in \mathcal{A}^k$.
The \Eq{gtens} may be extended on arbitrary elements
of the algebras due to distributivity. It is also
called {\em the skew tensor product}
and denoted as $\hotimes$ \cite{Post}.
\end{definition}

Clifford algebras are $\Z_2$-graded due to
the decomposition into subspaces generated by 
products with odd and even number of generators
and so {\em the skew tensor product} may be
defined as well.

\subsection{Structure of $\bm{\Cl_+(3n)}$}
\label{Sec:StrCl3n}

The skew tensor product is important for the description 
of Clifford algebras, because  
$\Cl(n+m) \isom \Cl(n)\hotimes\Cl(m)$,
$\Cl_+(n+m) \isom \Cl_+(n)\hotimes\Cl_+(m)$
and for complex case 
$\Cl(n+m,\C) \isom \Cl(n,\C)\hotimes\Cl(m,\C)$ \cite{ClDir,Post}. 
The repetition of the skew tensor product 
may be used for construction of $\Cl_+(3n)$ from $\Cl_+(3)$
\begin{equation}
 \Cl_+(3n) \isom \underbrace{\Cl_+(3) \hotimes \cdots \hotimes \Cl_+(3)}_n.
 \label{Cl3prod}
\end{equation}
A similar equation could be also written for $\Cl(3n)$ and $\Cl(3)$.
The analogue of \Eq{oxM2C} is also relevant
\begin{equation}
 \Cl(2n,\C) \isom \underbrace{\Cl(2,\C) \hotimes \cdots \hotimes \Cl(2,\C)}_n.
 \label{Cl2Cprod}
\end{equation}

For both $\Cl(3n)$ and $\Cl_+(3n)$ generators 
$\e^{(l)}_\nu$ \Eq{esub} may be represented as
\begin{equation}
\e^{(l)}_\nu = 
 {\underbrace{\Id\hotimes\cdots\hotimes \Id}_{l-1}\,}\hotimes
  \e_\nu\hotimes\underbrace{\Id\hotimes\cdots\hotimes\Id}_{n-l}.
\label{ehot}  
\end{equation}

Instead of \Eq{bip} for $\bi^{(k)}$ may be used 
\begin{equation}
\bi^{(k)} =
{\underbrace{\Id\hotimes\cdots\hotimes \Id}_{k-1}\,}\hotimes
  \bi\hotimes\underbrace{\Id\hotimes\cdots\hotimes\Id}_{n-k}.
\label{bipk}  
\end{equation}

It was already mentioned that the $Spin(3n)$ group 
used to be represented via $\Cl^0(3n)$,
but an equivalent construction with $\Cl_+^0(3n)$ 
may be more desirable here due to relation of
$\Cl_+(3)$ with the Pauli algebra \Eq{Cl3p}.

The representation of $Spin(3n)$ with the skew tensor product
justifies the idea of `$Spin(3n)$ circuits', because it has
more direct analogy with usual quantum circuits than
a rather formal $3{\times}n$ subdivision 
\Eq{esub} in the introduction, \Sec{intro}. 

Any quantum circuit with $n$ qubits corresponds
to an element of $SU(2^n)$ or some 
$2^n \times 2^n$ complex matrix from $M(2^n,\C)$
represented as the complex tensor product of $n$ 
Pauli algebras $M(2,\C)$ \Eq{oxM2C}.

Due to the isomorphism \cite{ClDir,Port} 
$M(2^n,\C) \isom \Cl(2n,\C)$ and \Eq{Cl2Cprod}
the complex $\Z_2$-graded tensor product may be used as well,
but the situation is more difficult for real algebras.

It was mentioned $M(2,\C)_\Re \isom \Cl_+(3)$ \Eq{CxH},
but the usual tensor product of $n$ copies of $\Cl_+(3)$
has the real dimension $2^{3n}$, and it may be `aggregated' 
into $M(2^n,\C)_\Re$ \Eq{MCHn} with the application
of {\em the correlator} $P_\iota$ \Eq{Pc}. 

The $\Cl_+(3n)$ may be represented as {\em the skew
tensor product} of $n$ algebras $\Cl_+(3)$ \Eq{Cl3prod}
with the real dimension $2^{3n}$, but the structure is
more complicated, because the `imaginary units' $\bi^{(k)}$ 
Eqs.~(\ref{bip}, \ref{bipk}) with different $k$ 
anticommute.

Now \Eq{zxbiy} and \Eq{CxH} should be applied to the skew 
tensor product \Eq{Cl3prod} and any $\bm{l} \in \Cl_+(3n)$ 
may be expressed as a composition
\begin{equation}
 \bm{l} = \sum_K \bm{c}_K \bm{h}_K, 
 \label{cKhK}
\end{equation}
where $\bm{c}_K$ are products of $\bi^{(k)}$, generating 
$2^n$-dimensional (sub)algebra isomorphic with $\Cl(n)$ 
and $\bm{h}_K$ are elements of $2^{2n}$-dimensional (sub)algebra
\begin{equation}
  \underbrace{\Cl^0_+(3) \hotimes \cdots \hotimes \Cl^0_+(3)}_n
  \isom \Q^{\otimes n}.
 \label{Hoxn} 
\end{equation}
In \Eq{Hoxn} a special symbol for the skew tensor product
is redundant, because for even subalgebras 
it coincides with the usual tensor product due to \Eq{gtens}.

All $\bm{c}_K$ in \Eq{cKhK} are products of $\bi^{(k)}$ and
commute with $\bm{h}_K$ from \Eq{Hoxn}. So, for any
$\bm{l},\bm{l}' \in \Cl_+(3n)$ 
\begin{equation}
 \bm{l}\bm{l}' 
 = \sum_K \bm{c}_K \bm{h}_K \sum_J \bm{c}'_J \bm{h}'_J
 = \sum_{K,J} (\bm{c}_K\bm{c}'_J) (\bm{h}_K\bm{h}'_J).
 \label{chKchJ}
\end{equation}
Due to \Eq{chKchJ} the decomposition \Eq{cKhK} satisfies
the formal definition of the tensor product of algebras 
\cite{Post,VDW} and so
\begin{equation}
 \Cl_+(3n) \isom \Cl(n) \otimes \Q^{\otimes n}.
 \label{ClHn}
\end{equation}

The direct consequence of the same constructions is
\begin{equation}
 \Cl_+^0(3n) \isom \Cl^0(n) \otimes \Q^{\otimes n},
 \label{Cl0Hn}
\end{equation} 
because the term $\Q^{\otimes n}$ \Eq{Hoxn} 
belongs to the even subalgebra and so 
the number of multipliers 
$\bi^{(k)} = \e_{123}^{(k)}$ should be also 
even in $\Cl_+^0(3n)$. 

\smallskip

The quantum circuits with $n$ qubits are described by 
the group $SU(2^n) \subset M(2^n,\C)$ and 
the resembling expression \Eq{MCHn} for $M(2^n,\C)_\Re$
is useful for the comparison with the structure of $\Cl_+(3n)$.

\subsection{Spin(6) group}
\label{Sec:Spin6}

Due to \Eq{Cl0Hn}
\begin{equation}
 \Cl^0_+(6) \isom \Cl^0(2) \otimes \Q^{\otimes 2}
 \isom \C_\Re \otimes \Q \otimes \Q 
 \isom M(4,\C)_\Re,
 \label{Cl06M4}
\end{equation}
where $\Cl^0(2) \isom \C_\Re$ with respect to
the `imaginary unit'
\begin{equation}
\di =  \bi^{(1\,2)} \equiv \bi^{(1)}\bi^{(2)},\quad
\di^2 = -\Id.
\label{i1i2}
\end{equation}
In fact, $\di$ is the product of all six
generators of the Clifford algebra and it commutes
with elements of the {\em even} subalgebra. 

Let us illustrate the isomorphism $Spin(6) \isom SU(4)$ 
already mentioned in the introduction, \Sec{intro}. It is convenient
to use the Lie algebra $spin(6)$ for the description
of the structure of the groups.
The basis of the Lie algebra $spin(n)$ ---
are products of pairs of generators \cite{ClDir}.

The basis of $spin(6)$ includes fifteen such pairs.
The six products 
$\e^{(l)}_{jk}$, $1 < j < k < 3$, $l=1,2$ 
are corresponding to the couple of different $spin(3)$ subalgebras. 
The structure of $Spin(3)$ was already discussed in \Sec{Spin3}
and with the notation used there in \Eq{es} the six elements
may be rewritten as $\es_j^{(l)}$, $j = 1,2,3$, $l=1,2$.

Other nine products 
$\e^{(1)}_j\e^{(2)}_k$, $j,k = 1,2,3$
may be rewritten as
$\di\es_j^{(1)}\es_k^{(2)}$,
$j,k = 1,2,3$, where
an `imaginary unit' $\di = \bi^{(1\,2)}$ 
was already introduced above \Eq{i1i2}. 

Let us show $spin(6) \isom su(4)$. The basis of the Lie algebra $su(4)$
--- are anti-Hermitian $4 \times 4$ matrices
(see {\em Note} in \Sec{Spin3}). Let us use for such a
purpose the representation with the tensor products of Pauli matrices.
The isomorphism may be directly shown with the map from
the basis of $spin(6)$ into anti-Hermitian matrices
\begin{subequations}
\label{s6su4}
\begin{align}
 \es_j^{(1)} \mapsto i\sigma_j \otimes \Id,\quad
 \es_j^{(2)} \mapsto \Id \otimes i\sigma_j\quad& (j = 1,2,3),
 \label{s6su4v1} \\ 
 \di\es_j^{(1)}\es_k^{(2)} \mapsto i\sigma_j \otimes \sigma_k
 \quad& (j,k = 1,2,3). \label{s6su4v2}
\end{align}
\end{subequations}
\Eq{s6su4v1} corresponds to \Eq{es}
for two $spin(3)$ subalgebras and \Eq{s6su4v2} represents
an {\em `entanglement'}.

\subsection{Spin(3n) circuits}
\label{Sec:Circ3n}

Let us consider decomposition of the $Spin(3n)$ group.
The basis of the Lie algebra $spin(3n)$ may be represented as
\begin{equation}
\e^{(l)}_j\e^{(m)}_k = \e_{3(l-1)+j}\e_{3(m-1)+k}.
\label{eljmk}
\end{equation}
An element of the Lie group $Spin(3n)$ may be expressed as
an exponent of the linear combination of \Eq{eljmk} or 
composed from a product of
\begin{equation}
U_\varepsilon^{(lm)} = \exp(\varepsilon\e^{(l)}_j\e^{(m)}_k).
\label{Ueps}
\end{equation}
Such approach is well known in the Lie-algebraic description
of (non)universal sets of quantum 
gates \cite{Vla0,DeuUn,DiVin95,SBOK6}.

For fixed $l \neq m $ \Eq{Ueps} describes elements from a subgroup 
of two-line gates $\mathcal{G}^{(lm)} \isom Spin(6)$.
The elements \Eq{Ueps} with $l=m$ are from a subgroup of 
one-line gates $\mathcal{G}^{(l)} \isom Spin(3)$. 
The one- and two-line gates are enough for construction of
the group $Spin(3n)$, because the basis of the Lie algebra $spin(3n)$ 
includes only the terms such as \Eq{eljmk}.

A more general analogue of $n$-line gates for $n > 2$ for $\Cl_+(3n)$
would include some subgroup of invertible elements of the 
algebra. In fact, the group of usual quantum $n$-qubit gates $SU(2^n)$ 
is isomorphic with a subgroup of such a group, because $M(2^n,\C)$ 
may be considered as a subalgebra of $\Cl_+(3n)$ due to
\Eq{MCHn} together with \Eq{ClHn}  
and an inclusion $\C_\Re=\Cl(1) \subset \Cl(n)$.

On the other hand, 
any $n$-line gate composed from one- and two-line gates
due to the structure of $\Cl_+(3n)$ is from a subgroup isomorphic with 
$Spin(3n)$. Let us compare dimensions of the groups: 
\begin{equation}
\dim SU(2^n) = 4^n-1,\quad
\dim Spin(3n) = \frac{3n(3n-1)}{2}.
\label{dims}
\end{equation}

The dimensions are not equal for $n > 2$ and due to \Eq{dims}
for $Spin(3n)$ the growth is quadratic with respect to $n$,
versus the exponential one for quantum circuits.
The comparison of dimensions for $n=1,\ldots,5$ is represented
in the table below.
\begin{equation}
\begin{array}{|c|r|r|r|r|r|}\hline
 n & 1 & 2 & 3 & 4 & 5 \\ \hline
 \dim SU(2^n)  & 3 & 15 & 63 & 255 & 1023 \\ \hline
 \dim Spin(3n) & 3 & 15 & 36 & 66  & 105 \\ \hline
\end{array}
\label{tabdims}
\end{equation}

The case $n=3$ may be considered for the illustration of a 
difference for the compositions of two-line gates.
It is convenient to use Lie algebras, because the structure
of products of elements of a Lie group is clear
from the bracket operation \Eq{comm} \cite{Vla0,DeuUn,DiVin95}.

Let us compare $spin(9)$ and the Lie algebra $su(8)$
of the Lie group $SU(2^3)$ of quantum three-qubit gates.
The Lie algebra $su(8)$ may be again represented
using tensor products with Pauli matrices to
comparison with analogues of \Eq{s6su4}.

Let us use notation
\begin{equation}
 \sigma_{j;l} = {\underbrace{\Id\otimes\cdots\otimes \Id}_{l-1}\,}\otimes
  \sigma_j\otimes\underbrace{\Id\otimes\cdots\otimes\Id}_{n-l}
\label{sigjl}  
\end{equation}

\Eq{s6su4v1} describes one-line gates
and only consideration of \Eq{s6su4v2} is not trivial.
So, it is necessary to consider elements of $su(8)$
such as
\begin{equation}
 i\sigma_{j;1} \sigma_{k;2}, \quad
 i\sigma_{j';2}\sigma_{k';3}
 \quad (j,k,j',k' = 1,2,3).
\label{su4su8} 
\end{equation}

An analogue of \Eq{su4su8} for inclusion of two copies of 
$spin(6)$ into $spin(9)$ may be written as 
\begin{equation}
\di\es_j^{(1)}\es_k^{(2)}, \quad 
\di'\es_{j'}^{(2)}\es_{k'}^{(3)}
 \quad (j,k,j',k' = 1,2,3),
\label{s6s9}  
\end{equation}
where $\di'$ is introduced below in \Eq{didi}.
Despite of the isomorphism $su(4) \isom spin(6)$ expressed 
by \Eq{s6su4}, 
the essential difference between \Eq{s6s9} and  \Eq{su4su8} 
is the pair of anticommuting `imaginary units'
\begin{equation}
\di = \bi^{(1\,2)},\quad
\di' = \bi^{(2\,3)} \equiv \bi^{(2)}\bi^{(3)},\quad
\di\di' = - \di'\di.
\label{didi}
\end{equation} 
Structures of Lie brackets \Eq{comm} becomes different 
due to the anticommuting elements. 

For $su(8)$  the brackets of elements \Eq{su4su8} are zero 
iff $k=j'$, but for $k \ne j'$ the commutator generates an element
of {\em third} order $i\sigma_{j;1}\sigma_{l;2}\sigma_{k';3}$,
where $k \ne l \ne j'$.
Conversely, for $spin(9)$ the brackets of 
elements \Eq{s6s9} are zero iff $k \ne j'$, but for $k = j'$ 
the commutator is an element of the {\em second} order
$\bi^{(1\,3)}\es_j^{(1)}\es_{k'}^{(3)}$.

\smallskip

\Eq{su4su8} and relevant \Eq{s6s9} correspond 
to the consideration of two-line gates from $\mathcal{G}^{(1\,2)}$
and $\mathcal{G}^{(2\,3)}$, but taking 
into account $\mathcal{G}^{(1\,3)}$ or
arbitrary triple of indexes $\mathcal{G}^{(l m n)}$
may be performed in the similar way.

\section{Matrix tensor product representation}
\label{Sec:matr}

\subsection{Clifford algebras $\bm{\Cl(2n,\C)}$ and $\bm{\Cl(4n,\C)}$}
\label{Sec:Cl4nC}

The real Clifford algebras was revisited above in the \Sec{SPIN3N}. 
They have rather irregular structures and may be isomorphic with
algebras of real, complex, quaternionic matrices and
also with doubles of such algebras \cite{Port,Post}, 
{\em e.g.}, see \Eq{Cl3} for $\Cl(3) \isom {}^2\Q$.
It may be convenient to include them
as subalgebras 
into complex Clifford algebras with even dimensions $\Cl(2n,\C)$ 
isomorphic with algebra $M(2^n,\C)$ of $2^n \times 2^n$ complex 
matrices \cite{ClDir,Port}.

\smallskip

Generators of Clifford algebras $\Cl(2n,\C) \isom M(2^n,\C)$
in the Jordan-Wigner representation
may be expressed as the tensor products of 
the Pauli matrices \cite{ClDir,JW,Weyl} 
\begin{subequations}
\label{e2k}
\begin{eqnarray}
 \e_{2k-1} & = &
  i\,{\underbrace{\sigma_3\otimes\cdots\otimes \sigma_3}_{k-1}\,}\otimes
 \sigma_1\otimes\underbrace{\Id\otimes\cdots\otimes\Id}_{n-k} \, ,
 \label{e2k1}\\
 \e_{2k} & = &
 i\,{\underbrace{\sigma_3\otimes\cdots\otimes \sigma_3}_{k-1}\,}\otimes
 \sigma_2\otimes\underbrace{\Id\otimes\cdots\otimes\Id}_{n-k} \, ,
 \label{e2k2}
\end{eqnarray}
\end{subequations}
where $k = 1,\ldots,n$.

For construction of $\Cl(4n,\C)$ may be used an analogue of \Eq{e2k}
with Dirac $4 \times 4$ matrices 
\begin{equation}
 \gamma^0 = -i\bmat{0&\Id\\\Id&0},\quad
 \boldsymbol\gamma = -i\bmat{0&\boldsymbol\sigma\\-\boldsymbol\sigma&0},\quad
 \gamma_5 = -i \gamma^0\gamma^1\gamma^2\gamma^3
  = \bmat{\Id&0\\0&-\Id},
\label{gams}
\end{equation}
where $\boldsymbol\sigma$, $\boldsymbol\gamma$ 
denote $\sigma_j$, $\gamma^j$ with $j=1,2,3$ \cite{Wein}.

Let us also rewrite \Eq{gams} with tensor products
of Pauli matrices
\begin{equation}
 \gamma^0 = -i\sigma_1 \otimes \Id,\quad
 \boldsymbol\gamma = \sigma_2 \otimes \boldsymbol\sigma, \quad
 \gamma_5 = \sigma_3 \otimes \Id.
\label{gamtens} 
\end{equation}

Let us consider
\begin{equation}
\e^{[k]}_j  =
  {\underbrace{\gamma_5\otimes\cdots\otimes \gamma_5}_{k-1}\,}\otimes
 \gamma^j\otimes\underbrace{\Id\otimes\cdots\otimes\Id}_{n-k} \, ,
\label{e4k}
\end{equation}
where $k = 1,\ldots,n$, $j = 0,\ldots,3$.

The elements \Eq{e4k} are anticommutative and they define 
a Clifford algebra with $\bigl(\e^{[k]}_0\bigr)^2=-1$ and
$\bigl(\e^{[k]}_j\bigr)^2 = 1$, $j \neq 0$.
The generators $\e^{[k]}_0$ and $i\e^{[k]}_j$ ($j = 1,2,3$)
may be used for construction of $\Cl(4n,\C) \isom M(4^n,\C)$.

\Eq{e4k} may be also rewritten using decompositions \Eq{gamtens}
with Pauli matrices and notation \Eq{sigjl}
\begin{subequations}
\label{e13k}
\begin{align}
\e^{[k]}_0 &= 
-i\Bigl(\prod_{l=1}^{k-1}\sigma_{3;2l-1}\Bigr)\,\sigma_{1;2k-1}
\label{e13k1}\\
\e^{[k]}_j &= 
\Bigl(\prod_{l=1}^{k-1}\sigma_{3;2l-1}\Bigr)\,\sigma_{2;2k-1}\sigma_{j;2k}
\quad (j =1,2,3).\label{e13k3}
\end{align}
\end{subequations}

The generators \Eq{e4k} corresponds to the {\em universal} Clifford 
algebra $\Cl(4n,\C)$ with dimension $2^{4n}$, because they
generate the complete basis of the algebra $M(4^n,\C)$ of 
$4^n \times 4^n$ complex matrices. Indeed, the Dirac matrices 
$\gamma^j$, $j = 0,\ldots,3$ together with products may be used
as the basis of $M(4,\C)$ and so the basis of $M(4^n,\C)$ 
may be constructed from $2^{4n}$ different products 
of elements
\begin{equation}
\gamma^{j;k}  =
  {\underbrace{\Id\otimes\cdots\otimes \Id}_{k-1}\,}\otimes
 \gamma^j\otimes\underbrace{\Id\otimes\cdots\otimes\Id}_{n-k} \, ,
\label{g4k}
\end{equation}
with $k = 1,\ldots,n$ and $j = 0,\ldots,3$.
Any element $\gamma^{j;k}$ may be expressed as 
a product of $\e^{[k]}_j$
\begin{equation} 
{\underbrace{\Id\otimes\cdots\otimes \Id}_{k-1}\,}\otimes
 \gamma_5\otimes\underbrace{\Id\otimes\cdots\otimes\Id}_{n-k} \, =
 -i \e^{[k]}_{0}\e^{[k]}_{1}\e^{[k]}_{2}\e^{[k]}_{3},\quad
\gamma^{j;k} = \e^{[k]}_j\prod_{l=1}^{k-1}%
\left(-i\e^{[l]}_{0}\e^{[l]}_{1}\e^{[l]}_{2}\e^{[l]}_{3}\right)
\label{ek2gk}
\end{equation}
and so, the basis of $M(2^{2n},\C) \isom \Cl(4n,\C)$ is also 
generated by $\e^{[k]}_j$ \Eq{e4k}.

\subsection{Clifford algebra $\bm{\Cl_+(3n)}$}
\label{Sec:Cl3in4C}

\begin{definition}
\label{defe3k}
Let us introduce $\Cl_+(3n)$ as subalgebra of 
$\Cl(4n,\C) \isom M(4^n,\C)$ with $3n$ generators 
$\e^{[k]}_j$ ($j = 1,2,3$ and $k = 1,\ldots,n$)
represented by \Eq{e4k} or \Eq{e13k3}.
\end{definition}

An analogue of \Eq{bip} may be written using \Eq{e13k3}
\begin{equation}
 \bi^{[k]} = \e^{[k]}_1\e^{[k]}_2\e^{[k]}_3 =
 i \Bigl(\prod_{l=1}^{k-1}\sigma_{3;2l-1}\Bigr)\,\sigma_{2;2k-1}.
 \label{mbip}
\end{equation}
Let us note, that all Pauli matrices in \Eq{mbip}
have {\em odd} positions in the decomposition \Eq{sigjl}.
On the other hand, \Eq{e13k3} consists of $\bi^{[k]}$ 
multiplied on a Pauli matrix in the {\em even} position    
\begin{equation}
\e^{[k]}_j = -i\bi^{[k]}\sigma_{j;2k}
\quad (j =1,2,3).
\label{e3kbi}
\end{equation}

Let us consider complex subalgebras $M_o$ and $M_e$ of
$M(4^n,\C)$ generated by products of elements \Eq{sigjl} with
Pauli matrices in {\em odd} and {\em even} positions respectively. 
Both subalgebras are isomorphic with $M(2^n,\C)$ represented
as the tensor products with $n$ complex $2 \times 2$ matrices.

Elements $\bi^{[k]}$ are anticommutative and 
generate a subalgebra of $M_o$ isomorphic with $\Cl(n)$.
With respects to isomorphisms 
$M_o \isom M(2^n,\C) \isom \Cl(2n,\C)$, 
elements $\bi^{[k]}$  \Eq{mbip} correspond to $n$
generators \Eq{e2k2}.

Such a decomposition of elements from $\Cl_+(3n)$ 
on $\Cl(n) \subset M_o$ and $M_e\isom M(2^n,\C)$ is a complex
analogue of \Eq{ClHn} from \Sec{StrCl3n}.

\subsection{Spin(3) and Spin(6) groups}
\label{Sec:MatSpin6}

The analogues of equations for $Spin(3)$ groups from \Sec{Spin3}
with generators $\e_j \equiv \e^{[1]}_j$ from
{\em definition \ref{defe3k}} are rather straightforward.
The products of two generators $\e_{jk}\equiv\e_j\e_k$ 
defined in \Eq{es}
may be represented using $4 \times 4$ matrices \Eq{gams} 
and tensor products \Eq{gamtens}
\begin{equation}
  \es_l = i\bmat{\sigma_l&0\\0&\sigma_j}
   = i\Id \otimes \sigma_l =i\sigma_{l;2}\quad (l=1,2,3).
  \label{mes} 
\end{equation}

\medskip 

Let us now represent the $Spin(6)$ group using
the Clifford algebra $\Cl_+(6)$.
In agreement with the {\em definition~\ref{defe3k}}
$$\Cl_+(6) \subset \Cl(8,\C) \isom M(16,\C)$$ and 
$\e^{[k]}_j$, $k=1,2$, $j = 1,2,3$  may be written as 
tensor products of two $4 \times 4$ matrices \Eq{e4k}
$$\e^{[1]}_j = \gamma^j \otimes \Id,\quad 
\e^{[2]}_j = \gamma_5 \otimes \gamma^j$$
and rewritten with Pauli
matrices using \Eq{e13k3}
$$\e^{[1]}_j = \sigma_{2;1}\sigma_{j;2},\quad
\e^{[2]}_j = \sigma_{3;1}\,\sigma_{2;3}\sigma_{j;4}.$$

Similarly with \Sec{Spin6} here is again convenient
to consider the Lie algebra $spin(6)$ with the basis defined by 
six products 
$\e^{[l]}_{jk}\equiv\e^{[l]}_j\e^{[l]}_k$, $1 < j < k < 3$, $l=1,2$ 
together with nine products $\e^{[1]}_j\e^{[2]}_k$, $j,k = 1,2,3$. 

The basis may be written down using 
{\em definition \ref{defe3k}} and \Eq{sigjl}.
The first six products describe two $spin(3)$ subalgebras 
similarly with \Eq{mes} 
\begin{subequations}
\label{ms6}
\begin{equation}
\es_j^{[1]} = i\sigma_{j;2},\quad 
\es_j^{[2]} = i\sigma_{j;4},
\label{ms6v1}
\end{equation}
where $j = 1,2,3$.
Other nine products are 
\begin{equation}
\e^{[1]}_j\e^{[2]}_k 
 = i\sigma_{1;1}\sigma_{j;2}\,\sigma_{2;3}\sigma_{k;4}
 = -i\sigma_{1;1}\sigma_{2;3}\es_j^{[1]}\es_k^{[2]}, 
\label{ms6v2}
\end{equation}
\end{subequations}
where $j,k = 1,2,3$.
The structure of \Eq{ms6} resembles \Eq{s6su4} used in
\Sec{Spin6} to illustrate the isomorphism $Spin(6) \isom SU(4)$.
It is also revised below in the \Sec{QCs6su4}.

\subsection{Spin(3n) group}
\label{Sec:MatSpin3n}

The group $Spin(3n)$ may be constructed with the method already
discussed in \Sec{Circ3n}. The Lie algebra $spin(3n)$ is
used for construction of the group. The basis of $spin(3n)$
is $\e^{[l]}_j\e^{[m]}_k$, $j,k = 1,2,3$.
 
The matrices $\e^{[l]}_j$, $j = 1,2,3$ in representations
\Eq{e4k} or \Eq{e13k3} are Hermitian, but products of
such elements are anti-Hermitian matrices, 
$$(\e^{[l]}_j\e^{[m]}_k)^\dag = \e^{[m]\dag}_k e^{[l]\dag}_j
= \e^{[m]}_k e^{[l]}_j = -\e^{[l]}_j\e^{[m]}_k.$$

Any element of $Spin(3n)$ is the exponent
of the linear combination of $\e^{[l]}_j\e^{[m]}_k$. 
Such an exponent of an anti-Hermitian 
matrix is {\em unitary}. 

It was already mentioned in \Sec{Circ3n},
any composition of one- and two-line gates 
\begin{equation}
U_\varepsilon^{[lm]} = \exp(\varepsilon\e^{[l]}_j\e^{[m]}_k).
\label{MUeps}
\end{equation}
is an element of the group $Spin(3n)$.
It may be more common for physical applications to introduce
Hamiltonians
\begin{equation}
 H^{[lm]}_{jk} = i\e^{[l]}_j\e^{[m]}_k
 \label{Hee}
\end{equation} 
and rewrite \Eq{MUeps} as
\begin{equation}
U_\tau^{[lm]} = \exp(-i\tau H^{[lm]}_{jk}).
\label{Utau}
\end{equation}

\section{Quantum circuits representation}
\label{Sec:QCirc}

\subsection{Quantum circuits model of Spin(3n)}
\label{Sec:QCs3n}

The quantum circuits framework for $\Cl_+(3n)$ and $Spin(3n)$
may be derived from the matrix tensor product 
representation discussed in the \Sec{matr}. 
Such a circuit model may be used both for the definition
of gates and {\em states}. 

The construction of a quantum circuit with $2n$ qubits
for modeling of the group $Spin(3n)$ using 
$\Cl(4n,\C) \isom M(4^n,\C)$ is rather straightforward.
It was already shown in \Sec{MatSpin3n} that 
all elements of the group $Spin(3n)$ are unitary matrices
and so it is a subgroup of the group $SU(2^{2n})$
of quantum gates with $2n$ qubits.

\smallskip

The decomposition on $M_o$ and $M_e$ mentioned
in \Sec{Cl3in4C} corresponds to partitions with $n$ 
qubits in odd and even position respectively.
It may be convenient sometimes to reorder qubits 
into even and odd subsystems using rearrangement
\begin{equation}
 (1,2,\ldots,2n) \mapsto
 (2,4,\ldots,2n),(1,3,\ldots,2n-1).
\label{treo} 
\end{equation}

\subsection{Isomorphism of Spin(6) and SU(4)}
\label{Sec:QCs6su4}

The structure of the $Spin(6)$ group may be modeled using 
a quantum circuit with four qubits. The element of $Spin(6)$ 
may be represented as the exponent of the linear combination of
matrices \Eq{ms6}. Let us write \Eq{ms6} for Hamiltonians
of quantum gates \Eq{Hee}

\begin{subequations}
\label{Hs6}
\begin{align}
& H_j^{[1]} 
  = \Id \otimes \sigma_j  \otimes \Id  \otimes \Id 
  = \sigma_{j;2},\qquad
 H_j^{[2]} 
  = \Id \otimes \Id  \otimes \Id  \otimes \sigma_j 
  = \sigma_{j;4},
\label{Hs6v1}\\
& H^{[12]}_{jk} 
  = \sigma_1 \otimes \sigma_j  \otimes \sigma_2  \otimes \sigma_k 
  = \sigma_{1;1}\sigma_{j;2}\,\sigma_{2;3}\sigma_{k;4}&
\label{Hs6v2}
\end{align}
\end{subequations}
and rewrite that
for reordering of qubits into even and odd subsystems \Eq{treo}  
\begin{subequations}
\label{Hs6eo}
\begin{align}
& H_j^{[1]} = \sigma_j \otimes \Id \otimes \Id \otimes \Id, \qquad 
 H_j^{[2]} = \Id \otimes \sigma_j\otimes \Id \otimes \Id,   
\label{Hs6eo1}\\
& H^{[12]}_{jk} 
  = \sigma_j\otimes \sigma_k \otimes \sigma_1 \otimes \sigma_2.   
\label{Hs6eo2}
\end{align}
\end{subequations}

An arbitrary Hamiltonian $H$ representing $Spin(6)$
is the linear combination of $15$
terms \Eq{Hs6eo} with real coefficients. Let us
write $H = H_1 + H_2$, with $H_1$ and $H_2$ are
corresponding to \Eq{Hs6eo1} and \Eq{Hs6eo2} respectively,
then
\begin{equation}
H_1 = H'_1 \otimes \Id \otimes \Id, \qquad
H_2 = H'_2 \otimes \sigma_1 \otimes \sigma_2,
\label{H12h12}
\end{equation}
where $H'_1$ and $H'_2$ are two-qubits Hamiltonians
(on the even subsystem). 

Let us now compare \Eq{Hs6eo} with
a basis of Hamiltonians for a system with two qubits
\begin{equation}
H'_{j;1} = \sigma_j \otimes \Id , \quad 
H'_{j;2} = \Id \otimes \sigma_j, \quad   
H'_{jk} = \sigma_j\otimes \sigma_k  
\label{Hsu4}.
\end{equation}
The Hamiltonians \Eq{Hsu4} coincide with the first two terms
in \Eq{Hs6eo}. The decomposition $H = H_1+H_2$ together
with \Eq{H12h12} may be used to define the Hamiltonian 
$H' = H'_1+H'_2$ on a system with two qubits. Due to one-to-one
correspondence between \Eq{Hsu4} and \Eq{Hs6eo},
any $H'$ may be constructed in such a way from some $H$
and vice versa.

Let us consider action of Hamiltonians \Eq{Hs6eo} on a system
of four qubits decomposed into even and odd subsystems
$\ket{\Psi_e}\ket{\Upsilon_o}$, there $\ket{\Psi_e}$ is
an arbitrary state of two qubits and 
$\ket{\Upsilon_o}$ is an eigenstate with unit eigenvalue of
the operator $\sigma_1 \otimes \sigma_2$. Composing eigenvectors of 
$\sigma_1$, $\sigma_2$ with equal eigenvalues
$\pm1$ it may be obtained
\[
 \ket{\Upsilon_o^{++}} = 
 \frac{1}{2}\bigl(\ket{0}+\ket{1}\bigr)\bigl(\ket{0}+i\ket{1}\bigr),
 \quad 
 \ket{\Upsilon_o^{--}} = 
 \frac{1}{2}\bigl(\ket{0}-\ket{1}\bigr)\bigl(\ket{0}-i\ket{1}\bigr).
\]
A linear combination of the states also may be used
\begin{equation}
\ket{\Upsilon_o} = \alpha \ket{\Upsilon_o^{++}}
+\beta\ket{\Upsilon_o^{--}},\quad 
|\alpha|^2+|\beta|^2 = 1.
\label{Ups}
\end{equation}

Let us consider the Hamiltonian
$H = H_1 + H_2$ introduced above with \Eq{H12h12}, then
\begin{equation}
H_1\bigl(\ket{\Psi_e}\ket{\Upsilon_o}\bigr) + 
H_2\bigl(\ket{\Psi_e}\ket{\Upsilon_o}\bigr)
= \bigl(H'\ket{\Psi_e}\bigr)\ket{\Upsilon_o}.
\label{HPsiUps}
\end{equation}
where $H' = H'_1+H'_2$ is the two-qubit Hamiltonian
also defined earlier using \Eq{H12h12}.
It may be derived directly  from \Eq{HPsiUps}, that 
a quantum gate corresponding to $Spin(6)$ for such 
a state acts as usual quantum gate on the first two qubits:
\begin{equation}
 e^{-i H \tau}\bigl(\ket{\Psi_e}\ket{\Upsilon_o}\bigr) = 
 \bigl(e^{-i H' \tau}\ket{\Psi_e}\bigr)\ket{\Upsilon_o}.
\label{UPsiUps} 
\end{equation}
It ensures one-to-one correspondence
$SU(4) \isom Spin(6)$ between the arbitrary gate
on two qubits and the $Spin(6)$ gate.

\subsection{Decomposition of Spin(3n)}
\label{Sec:QCdec3n}

It was already discussed in \Sec{Circ3n} that
despite of isomorphism of $Spin(6)$ with group
$SU(4)$ of two-qubit gates,  $Spin(3n)$ circuits 
composed from such $Spin(6)$ gates may have only 
quadratic dimension with respect to $n$.

An analysis with the quantum circuits model is
very similar.
In simplest case $Spin(9)$ group may be represented by 
six qubits reordered into even and odd subsystems
with three qubits in each using \Eq{treo}.
The Hamiltonians corresponding to overlapped
two-line gates \Eq{s6s9} in \Sec{Circ3n} may be constructed 
using \Eq{Hs6eo2}
\begin{equation}
\begin{split}
 H^{[12]}_{jk} 
 &= (\sigma_j\otimes \,\sigma_k\, \otimes \Id\,) \otimes 
 (\sigma_1 \otimes \sigma_2 \otimes \Id),\\
 H^{[23]}_{j'k'} 
 &= (\Id \otimes \sigma_{j'}\otimes \sigma_{k'}) \otimes  
    (\Id \otimes \sigma_1 \otimes \sigma_2).    
\label{H123eo}
\end{split}
\end{equation}

Lie-algebraic approach 
\cite{Vla0,DeuUn,DiVin95,SBOK6} 
uses Hamiltonians together with all possible 
commutators for analysis of quantum circuits. 

Due to \Eq{H123eo} the commutator of
$H^{[12]}_{jk}$, $H^{[23]}_{j'k'}$ is nonzero
only for $k = j'$ and produces Hamiltonian
of `second order'
\begin{equation}
 H^{[13]}_{jk'} = i\e^{[1]}_j\e^{[3]}_{k'} 
   = (\sigma_j \otimes \Id \otimes  \sigma_{k'}) \otimes  
     (\sigma_1 \otimes \sigma_3 \otimes \sigma_2).
\label{H13eo}      
\end{equation}

In more general case the situation is similar and 
the group $Spin(3n)$ is generated by Hamiltonians \Eq{Hee} with
two Pauli matrices in the {\em `primary'} (even) subsystem.
The number of Pauli matrices in the {\em `auxiliary'} (odd) subsystem
for $H^{[lm]}_{jk}$ ($l < m$) is $m-l+1$. Such a Hamiltonian
has structure resembling \Eq{H13eo} with $m-l-1$ matrices $\sigma_3$
inserted between $\sigma_1$ and $\sigma_2$. 

\section{Classical simulation}
\label{Sec:ClSim}
\subsection{General methods}
\label{Sec:ClSimGen}

An idea of {\em the classical simulation} of the $Spin(3n)$
circuit used here is analogous with the approach
used in \cite{Joz8,Joz9} for $\Cl(2n)$ and $Spin(2n)$. 

Few distinctions between `$Spin(3n)$ circuits'
and models related with $Cl(2n)$ may be analyzed
using $3n$ generators $\e_j^{[k]}$ 
of $\Cl_+(3n)$ defined by \Eq{e4k} or \Eq{e13k3} and 
$2n$ generators $\e_k$ of $\Cl(2n)$ from \Eq{e2k}.

%
The product $i\e_{2k-1}\e_{2k} = \sigma_{3;k}$ 
(denoted in \cite{Joz8,Joz9} as $Z_k$) is
an action of $\sigma_3$ on the qubit with index $k$.
It may be compared with 
$\sigma_{3;2k} = i\e_1^{[k]}\e_2^{[k]}$, where
an even index $2k$ corresponds to the initial 
order of qubits {\em without} any reordering.
Other Pauli matrices for qubits with even
indexes may be expressed as well:
$\sigma_{1;2k} = i\e_2^{[k]}\e_3^{[k]}$,
$\sigma_{2;2k} = i\e_3^{[k]}\e_1^{[k]}$.
An expression for qubits with odd indexes
is different and may be written using $\gamma_5$
from \Eq{gamtens}
$\sigma_{3;2k-1} = -i\e_0^{[k]}\e_1^{[k]}\e_2^{[k]}\e_3^{[k]}$.

Due to such a property the variation of setup used
in \cite{Joz8,Joz9} is:
\textit{ 
\begin{enumerate}
 \item the `$Spin(3n)$ circuit' with $2n$ qubits  
 \item the input state is any product state 
 \item the output is a measurement of a single qubit: 
   \begin{enumerate} 
    \item arbitrary for even indexes 
    \item in the computational basis for odd indexes 
    \label{oddline}
   \end{enumerate}  
\end{enumerate}}

Let us first consider qubits {\em with even indexes}.
Without the lost of generality {\em only measurements in 
computational basis} may be discussed, because any
one-qubit gate may be implemented on the even qubit and
it may be used for a measurement in another basis.

For such a simplified case methods from \cite{Joz8,Joz9}
may be applied with minimal modifications.
Let us consider $U= U_1 U_2 \cdots U_N$ representing 
element of $Spin(3n)$ group as a circuit
with $N$ gates
\begin{equation}
 \ket{\Psi_U} = U \ket{\Psi} 
  = U_1 U_2 \cdots U_N \ket{\Psi}.
 \label{PsiU}
\end{equation}
If $p_0^{(2k)}$ and $p_1^{(2k)}$ are probabilities of outcomes
of measurements in the computational basis for a qubit
with an index $2k$
\begin{equation}
 p_0^{(2k)} - p_1^{(2k)} = \bra{\Psi_U}\sigma_{3;2k}\ket{\Psi_U}
 = \bra{\Psi}U^\dag i\e_1^{[k]}\e_2^{[k]} U\ket{\Psi}.
\label{p0m1ev} 
\end{equation}

Let us note
$U^\dag \e_1^{[k]} \e_2^{[k]} U = 
U^\dag \e_1^{[k]}U\,U^\dag\e_2^{[k]} U$
and use a standard property of Spin groups \Eq{SpinSO},
rewritten as
\begin{equation}
U^\dag\e_j^{[k]}U = 
 \sum_{j',k'} R^{[kk']}_{jj'}\e_{j'}^{[k']},
\label{UejkU} 
\end{equation}
where $R^{[kk']}_{jj'}$ denotes
$3n \times 3n$ orthogonal matrix with elements
$R_{3(k-1)+j,3(k'-1)+j'}$. 

If the operator $U$ in \Eq{PsiU}  corresponds
to a decomposition of `$Spin(3n)$ circuits' on a sequence
of $N$ gates, the matrix 
also may be presented as a product $R = R_1 R_2 \cdots R_N$
with each term corresponding to a gate in the sequence
and it may be computed in time poly$(n,N)$. 

\Eq{p0m1ev} may be rewritten
\begin{align}
 p_0^{(2k)} - p_1^{(2k)} &= \bra{\Psi}i
  \Bigl(\sum_{j',k'} R^{[kk']}_{1j'}\e_{j'}^{[k']}\Bigr)
  \Bigl(\sum_{j',k'} R^{[kk']}_{2j'}\e_{j'}^{[k']}\Bigr)
  \ket{\Psi} \notag\\
  &= \sum_{j',k',j'',k''}\Bigl(R^{[kk']}_{1j'}R^{[kk'']}_{2j''}
    \bra{\Psi}i\e_{j'}^{[k']}\e_{j''}^{[k'']}\ket{\Psi}\Bigr).
 \label{RRPeeP}   
\end{align}
For the sum of $(3n)^2$ elements
$\bra{\Psi}i\e_{j'}^{[k']}\e_{j''}^{[k'']}\ket{\Psi}$ with
$\ket{\Psi} = \ket{\psi_1} \ldots \ket{\psi_{2n}}$ and 
{\em product} operators
\begin{equation} 
\e_{j'}^{[k']}\e_{j''}^{[k'']} =
{\mathbf{s}_1}^{k' k''}_{j'  j''}\! \otimes  
\cdots  \otimes\, {\mathbf{s}_{2n}}^{k' k''}_{j'  j''}
\label{eeprod}
\end{equation} 
each element is a product of $2n$
factors $\bra{\psi_m}{\mathbf{s}_{m}}^{k' k''}_{j'  j''}\ket{\psi_m}$,
$m = 1,\ldots,2n$.

Thus, the result of a measurement of a single qubit {\em with an
even index} may be computed in time poly$(n,N)$, where $n$ and $N$
are numbers of qubits and gates respectively.
Here the resemblance with the approach \cite{Joz8,Joz9}
to matchgates is quite clear. 

For qubits {\em with odd indexes} the difference is rather 
not essential.
Instead of \Eq{p0m1ev}, for a qubit with an index $2k-1$
should be used
\begin{equation}
 p_0^{(2k-1)} - p_1^{(2k-1)} = -\bra{\Psi}U^\dag 
 i\e_0^{[k]}\e_1^{[k]}\e_2^{[k]}\e_3^{[k]}U\ket{\Psi}.
\label{p0m1od} 
\end{equation}
Only $\e_0^{[k]}$ is not affected by the
$Spin(3n)$ group and modifications of 
other three elements $\e_j^{[k]}$ $(j = 1,2,3)$ 
are described by \Eq{UejkU}. 
Instead of \Eq{RRPeeP}, similar sum with $(3n)^3$ terms
should be written
\begin{equation}
 p_0^{(2k-1)} - p_1^{(2k-1)} =
 \sum_{\substack{k_1,k_2,k_3,\\j_1,j_2,j_3}}\!
 \Bigl(R^{[kk_1]}_{1j_1}R^{[kk_2]}_{2j_2}R^{[kk_3]}_{3j_3}
 \bra{\Psi}i\e_0^{[k]}
 \e_{j_1}^{[k_1]}\e_{j_2}^{[k_2]}\e_{j_3}^{[k_3]}\ket{\Psi}\Bigr).
 \label{RRRPeeeP}
\end{equation} 
For the product states $\ket{\Psi}$ and operators
$\e_0^{[k]}\e_{j_1}^{[k_1]}\e_{j_2}^{[k_2]}\e_{j_3}^{[k_3]}$
each $\bra{\Psi} \ldots \ket{\Psi}$
again may be expressed by multiplication
of $2n$ terms $\bra{\psi_m} \ldots \ket{\psi_m}$. 

The consideration shows, the result of a measurement 
in the computational basis of a single qubit {\em with odd index}  
may be again computed in time poly$(n,N)$. 

\subsection{Two-qubit gates}
\label{Sec:ClSimTwo}
\subsubsection{Hamiltonians}
\label{Sec:ClSimTwoH}

In the proof of effective classical simulation discussed 
in \Sec{ClSimTwo} was successfully used known approach \cite{Joz8,Joz9},
but the model itself has new properties. A promising 
achievement is possibility to implement arbitrary two-qubit gate.
Let us discuss that before consideration of general
case with many qubits.

Such an attainment in comparison with matchgate model is
associated with improved control over `primary' qubits with 
even indexes accompanied by separation of 
`auxiliary' qubits with odd indexes.
The example with two qubits implemented by circuit
with two additional auxiliary qubits was discussed in \Sec{QCs6su4}.
The model allows us to apply arbitrary unitary gate on pair
of qubit with fixing auxiliary qubits in undisturbed 
state $\ket{\Upsilon_o}$.

The Hamiltonians of two-qubit gates used in \Sec{QCs6su4}
are revisited here before discussion in \Sec{ClSimTwoU}
about possibility to work directly with unitary gates.
Let's use an opportunity to fix state $\ket{\Upsilon_o}$
of auxiliary qubits with odd indexes to focus on the qubits 
{\em with even indexes} and rearrange them \Eq{treo} to 
apply notation in agreement with expressions for two-qubit 
Hamiltonians such as \Eq{Hsu4}.

Let us rewrite \Eq{Hsu4} using \Eq{Hs6}
\begin{subequations}
\label{S6e2H}
\begin{align}
\es_j^{[1]} &\longleftrightarrow
-i H'_{j;1} = -i\sigma_j \otimes \Id , \qquad 
\es_j^{[2]} \longleftrightarrow
-i H'_{j;2} = -\Id \otimes i \sigma_j,
\label{S6e2H1}\\   
\e^{[1]}_j\e^{[2]}_k &\longleftrightarrow
-i H'_{jk} = -i \sigma_j\otimes \sigma_k,  
\label{S6e2H2}
\end{align}
\end{subequations}
where $\es_j^{[l]}$, $l=1,2$ are products of two generators 
$\e_p^{[l]}\e_k^{[l]}$, $j \ne p \ne k$ introduced 
in \Sec{Spin3}, \Eq{es} and reused in \Sec{MatSpin6}, \Eq{ms6v1}.
Thus, \Eq{S6e2H} describe $3+3+9=15$ different 
products with two generators of Clifford algebra $\Cl_+(6)$.

\Eq{S6e2H} may be also rewritten using notation 
$\sigma_j^{(1)} = \sigma_j \otimes \Id$,
$\sigma_j^{(2)} = \Id \otimes \sigma_j$. 
\begin{subequations}
\label{S6e2s}
\begin{align}
&\e^{[1]}_j\e^{[1]}_k  \longleftrightarrow
-\sigma_j^{(1)}\sigma_k^{(1)} , \qquad 
\e^{[2]}_j\e^{[2]}_k  \longleftrightarrow
-\sigma_j^{(2)}\sigma_k^{(2)},
\label{S6e2s1}\\   
&\e^{[1]}_j\e^{[2]}_k \longleftrightarrow
-i\sigma_j^{(1)}\sigma_k^{(2)}.  
\label{S6e2s2}
\end{align}
\end{subequations}
The imaginary unit multiplier presenting in \Eq{S6e2s2},
but missing in \Eq{S6e2s1} is important. Formally,
\Eq{S6e2s} are in complete agreement with \Eq{S6e2H} 
due to law of multiplication of Pauli matrices
and it illustrates the fact that such correspondence
is defined {\em only for pairs} of generators.
The $\e_j^{[l]}$ are operators on four qubits 
that should be rewritten after reordering \Eq{treo} as
\[
 \e_j^{[1]} = (\sigma_j \otimes \Id) \otimes (\sigma_2 \otimes \Id),\quad
 \e_j^{[2]} = (\Id \otimes \sigma_j) \otimes (\sigma_3 \otimes \sigma_2)
\]
and, so, the generators should not be confused with
elements such as $\sigma_j^{(l)}$.

It is clear from \Eq{S6e2H} that expression 
$\exp(-i H \tau)$ used for construction of unitary
gates contains only {\em real} coefficients 
after rewriting with products of two generators.
However, $4 \times 4$ unit matrix $\Id \otimes\Id$ is not presented 
in \Eq{S6e2H} or \Eq{S6e2s} and all fifteen matrices described by the 
equations are {\em traceless}. The traceless Hamiltonians 
corresponds to unitary matrices with unit determinant, 
{\em i.e.}, to the group $SU(4)$. 

\smallskip

Let's describe the method of construction of element 
of $Spin(6)$ group for a quantum gate with known representation
as $U = \exp (-i H \tau)$ with some real $\tau$ and Hermitian $H$.

The matrix $H$ may be considered traceless without lost of
generality by transition 
\[H \to H - \bigl(\Tr(H)/4\bigr)\,\Id\otimes\Id .\]
Let's enumerate Hamiltonians in \Eq{S6e2H} with single 
index: $H_J$, $J=1,\ldots,15$.
It may be simply checked that $\Tr(H_J H_K) = 4 \delta_{JK}$.
Due to that property coefficients of decomposition
of a traceless Hamiltonian $H$ using basis \Eq{S6e2H}
may be expressed as 
\begin{equation}
H = \sum_{J=1}^{15} h_J H_J,\quad
h_J = \Tr(H_J H)/4, 
\label{decompH15}
\end{equation}
where any index $J$ is associated with known product of two  
generators. It was mentioned earlier that 
such products correspond to Lie algebra of $spin(6)$
and so such algorithm produces element $\mathbf{s}_H \in spin(6)$
for any traceless Hamiltonian $H$.
Now, element $\mathbf{S}_H \in Spin(6)$ group may be
expressed as $\mathbf{S}_H = \exp(\mathbf{s}_H \tau)$.

\subsubsection{Gates}
\label{Sec:ClSimTwoU}

It may look more convenient instead of Hamiltonians 
used in \Sec{QCs6su4} and \Sec{ClSimTwoH}
to work directly with unitary gates. 
Sometimes exponential representation also can be useful,
{\em e.g.}, any operator $\mathfrak{J}$ with property 
$\mathfrak{J}^2=-\Id$ complies with a simple equation
\[
\exp(\tau\mathfrak{J}) =\cos(\tau)\Id+\sin(\tau)\mathfrak{J},
\]
and together with standard property of exponent for commuting 
operators $\mathfrak{J}$, $\mathfrak{K}$
\[
\exp(\mathfrak{J}+\mathfrak{K}) = \exp(\mathfrak{J})\exp(\mathfrak{K})
\]
it can be applied to some interesting examples. 

However, a method of direct construction of element 
$\mathbf{S} \in Spin(6)$ from a quantum gate $U \in SU(4)$
is required in more general case. The reason to avoid matrices from 
$U(4)$ with non-unit determinants was illustrated above 
in \Sec{ClSimTwoH} and to exploit isomorphism $Spin(6) \isom SU(4)$
a gate should be tuned using a phase multiplier: $U'=\det(U)^{-1/4} U $.

The unitary gate must be mapped into element of $Spin(6)$ group and
such element may contain products with any even number of generators.
Thus, it is necessary to consider 32 such products instead
of only 15 discussed below and already used in \Eq{decompH15}. 
Such consideration includes unit, product
of all six generators denoted earlier by $\di$ in \Eq{i1i2},
fifteen products with two generators already
used earlier and fifteen products with four generators.
Two last numbers are the same, because 
any product of four generators can be expressed as a
pair multiplied on $\di$.

In such a way the Spin(6) group is represented as some
subspace of algebra $\Cl^0_+(6)$ already discussed
in \Sec{Spin6}.
A possible confusion may
appear because the 32-dimensional {\em real} algebra $\Cl^0_+(6)$
used for construction of $Spin(6)$ should be mapped into algebra 
of all $4 \times 4$ {\em complex} matrices.

It was already discussed earlier in \Sec{Spin6} with
basic idea to use $\di^2=-1$ commuting with all elements of 
$\Cl^0_+(6)$ as an imaginary unit. Such general approach is in 
agreement with \Eq{S6e2H} of \Eq{S6e2s}, because
they lead to
\[
 \di = (\e^{[1]}_1\e^{[1]}_2) (\e^{[1]}_3\e^{[2]}_1)
 (\e^{[2]}_2\e^{[2]}_3) \leftrightarrow
 (-i\sigma_3\otimes\Id) (-i\sigma_3\otimes\sigma_1)
  (-\Id\otimes i\sigma_1) = i \Id \otimes \Id.
\]
Relation between products with four 
and two generators obtained by multiplication on $\di$ 
was already mentioned above. Thus, the fourfold products 
correspond to \Eq{S6e2H} {\em without imaginary units}.

Let's finally describe the algorithm for the map 
$U \in SU(4) \rightarrow Spin(6)$. The unit matrix
together with fifteen matrices \Eq{S6e2H} may be used
as a basis $H_J$, $J=0,\ldots,15$. Similarly
with \Eq{decompH15} for given
matrix $U \in SU(4)$ may be calculated sixteen
complex coefficients $u_J$
\begin{equation}
U = \sum_{J=0}^{15} u_J U_J,\quad
u_J = \Tr(H_J U)/4 
\label{decompU16}
\end{equation}
Real and imaginary parts of $u_0$ corresponds to 
unit and $\di$ respectively. For $J \ge 1$ real part
of $u_J$ is responsible for {\em the same pair} of generators 
as $h_J$ in \Eq{decompH15} and imaginary part of $u_J$ conforms to 
product of four generators obtained from this pair 
by multiplication on $\di$.

\bigskip

After construction of $S \in Spin(6)$ using methods
discussed above, matrix $R \in SO(6)$ is defined 
similarly with \Eq{UejkU}
\begin{equation}
S^{-1}\e_j^{[k]}S = 
 \sum_{j'=1}^3\sum_{k'=1}^2 R^{[kk']}_{jj'}\e_{j'}^{[k']}.
\label{SejkS} 
\end{equation}
The expression \Eq{v12} below for six-dimensional vector affected 
by such rotations makes more clear structure of double indexes 
used in \Eq{SejkS}
\begin{equation}
 \bm{v} = 
  v^{[1]}_1\e^{[1]}_1+v^{[1]}_2\e^{[1]}_2+v^{[1]}_3\e^{[1]}_3+
  v^{[2]}_1\e^{[2]}_1+v^{[2]}_2\e^{[2]}_2+v^{[2]}_3\e^{[2]}_3.
\label{v12}  
\end{equation}

For clarification, in Appendix A is presented a program 
for computer algebra system calculating elements of $Spin(6)$ 
group and rotation for given $4\times 4$ matrix of two-qubit gate.

Techniques developed here for six generators
$\e_{j}^{[1]}$ and $\e_{j}^{[2]}$ are simply generalized
for any pair $\e_{j}^{[k']}$ and $\e_{j}^{[k'']}$ 
and two-qubit gates on 
`primary' (even) indexes $2k'$ and $2k''$. 
Any two-qubit gate on given indexes is mapped into
rotation of 6D subspace similar with \Eq{v12}
\begin{equation}
 \bm{v} = 
  v^{[k']}_1\e^{[k']}_1+v^{[k']}_2\e^{[k']}_2+v^{[k']}_3\e^{[k']}_3+
  v^{[k'']}_1\e^{[k'']}_1+v^{[k'']}_2\e^{[k'']}_2+v^{[k'']}_3\e^{[k'']}_3.
\label{vk'k''}
\end{equation}

\subsection{Simulation of Spin(3n) circuits with suitable states}
\label{Sec:SimSuit}

The improved control over `primary' qubits with even 
indexes can be considered as essential contribution of presented model 
and it is acceptable for a while to avoid detailed
consideration of 
qubits with odd indexes \Eq{p0m1od} and \Eq{RRRPeeeP}.

The method to save states of the `auxiliary' qubits 
used above to exclude them from consideration 
does not work for $n > 2$. Let's consider \Eq{H123eo} for
illustration of the case $n=3$. Operators acting on three
auxiliary qubits for $H^{[12]}$ and $H^{[23]}$ are anticommute
in agreement with \Eq{didi} and common eigenstate 
$\ket{\Upsilon}$ (with nonzero eigenvalue) 
can not exist for them.

Thus, without requirement about specific states for qubits with odd indexes,
initial state may be chosen similarly with
matchgate circuits $\ket{\Psi_0} =\ket{00\ldots0}$ \cite{Joz9}.
Probabilities \Eq{RRPeeP} for given initial state 
are directly calculated using values
\begin{equation}
\mu^{[k'k'']}_{j'j''} 
 = \bra{\Psi_0}i\e_{j'}^{[k']}\e_{j''}^{[k'']}\ket{\Psi_0}. 
\label{mukkjj}
\end{equation}

For $k'<k''$ \Eq{g4k} together
with \Eq{gamtens} provide expressions such as
\[
i\e_{j'}^{[k']}\e_{j''}^{[k'']} =
\underbrace{\Id\otimes\cdots\otimes\Id}_{2(k'-1)\,}\otimes\, 
 \sigma_1\otimes\sigma_{j'}
 \otimes \underbrace{\sigma_3\otimes\Id\otimes\cdots
 \otimes\sigma_3\otimes\Id}_{2(k''-k'-1)}\otimes\,
 \sigma_2\otimes\sigma_{j''}\otimes
 \underbrace{\Id\otimes\cdots\otimes\Id}_{2(n-k'')}.
\]

\Eq{mukkjj} for the product states  includes factors 
$\bra{0}\sigma_1\ket{0}$ and $\bra{0}\sigma_2\ket{0}$ 
equal to zero and so for $k' \neq k''$, $\mu^{[k'k'']}_{j'j''} = 0$.
The factors are due to qubits with odd indexes and, moreover, 
the expression vanishes for initial states of `auxiliary' 
qubits either $\ket{0}$ or $\ket{1}$.
 
For $k' = k''$ 
\[
i\e_{j'}^{[k']}\e_{j''}^{[k']} =
\underbrace{\Id\otimes\cdots\otimes\Id}_{2(k'-1)\,}\otimes\, 
 \Id\otimes(i\sigma_{j'}\sigma_{j''})\otimes
 \underbrace{\Id\otimes\cdots\otimes\Id}_{2(n-k')}.
\]
Thus, $\mu^{[k'k']}_{j'j''} = \bra{0}i\sigma_{j'}\sigma_{j''}\ket{0}
= i\delta_{j'j''} -\epsilon_{j'j''3}$,
where $\epsilon_{abc}$ is totally antisymmetric Levi-Civita symbol 
(permutation tensor). 
Formal expression for both cases $k'=k''$ and $k'\neq k''$ may be
written as
\begin{equation}
\mu^{[k'k'']}_{j'j''} = 
(i\delta_{j'j''}-\epsilon_{j'j''3})\delta_{k'k''} 
\label{mukkjjs}
\end{equation} 
However, the term with imaginary unit in \Eq{mukkjjs} is redundant and
it does not produce any contribution in final expression \Eq{RRPeeP} 
because
\begin{equation}
 \sum_{j',k',j'',k''}\!
   R^{[kk']}_{1j'}R^{[kk'']}_{2j''}
    \delta_{j'j''}\delta_{k'k''} =
   \sum_{j',k'} R^{[kk']}_{1j'}R^{[kk']}_{2j'} = 0
 \label{RRzero}   
\end{equation}
Indeed, in `plain' notation
$R^{[kk']}_{jj'}=R_{3(k-1)+j,3(k'-1)+j'}$ is
$3n \times 3n$ orthogonal matrix and 
\Eq{RRzero} corresponds to scalar product of two different columns
\[
   \sum_{m=1}^{3n}R_{3(k-1)+1,m}R_{3(k-1)+2,m}.
\]
for the matrix of rotation $R$, but they are 
orthogonal and the scalar product is zero.

The term with $j'=j''$, $k'=k''$ corresponds 
to product of two equal generators 
$\e_{j}^{[k]}\e_{j}^{[k]}=\Id$ and could be
from very beginning excluded from consideration 
for any initial state $\ket{\Psi}$.

Without this vanished term \Eq{RRPeeP} may be rewritten for 
$\ket{\Psi_0}$
\begin{equation}
 p_0^{(2k)} - p_1^{(2k)} = -\sum_{\substack{k',k''\\j',j''}}
   R^{[kk']}_{1j'}R^{[kk'']}_{2j''}\epsilon_{j'j''3}\delta_{k'k''}
    = \sum_{k'=1}^n (R^{[k k']}_{12}R^{[k k']}_{21} 
                     - R^{[k k']}_{22}R^{[k k']}_{11}).
 \label{RRPeeP0}   
\end{equation}

Other initial states may be considered as
well, because any one-qubit transformation can be implemented 
for `primary' qubits
by $Spin(3n)$ circuit and used for altering $\ket{\Psi_0}$ in
a desired way. The derivation above shows also 
that the \Eq{RRPeeP0} is valid only for initial states of
`auxiliary' qubits either $\ket{0}$ or $\ket{1}$.

Techniques developed in \Sec{ClSimTwo} work for six generators
$\e_{j}^{[k']}$ and $\e_{j}^{[k'']}$ with any pair of indexes 
$k'$ and $k''$ and provide method to map any two-qubit gate into 
6D rotation on subspace \Eq{vk'k''}.
It is true for arbitrary states of `auxiliary' qubits, yet
$\ket{\Upsilon_o}$ were used earlier for convenience
to avoid entanglement between `primary' and `auxiliary' qubits. 

In presented method of simulation the final matrix $R$ is 
obtained as composition of rotations on the
6D subspaces and with the known matrix $R$ probabilities
for measurement of `primary' qubits are calculated using 
\Eq{RRPeeP0}. Exertion of \Eq{RRRPeeeP} for `auxiliary' 
qubits with chosen initial states is not discussed in 
presented work.

\section{Conclusions}
\label{Sec:Concl}

This work was devoted to relations between
$Spin(3n)$ groups and quantum circuits.
Any transformation of one or two qubits may be described by 
$Spin(3)$ or $Spin(6)$ groups respectively. 
Such a property together with well-known correspondence 
between group $Spin(2n)$ and some non-universal 
quantum circuits with $n$ qubits causes a natural 
question about similar relations for $Spin(3n)$ groups.

The connection between
quantum circuits with $n$ qubits and $Spin(3n)$ may
be illustrated using the Clifford algebra $\Cl_+(3)$
isomorphic with the Pauli algebra. 
On the one hand, {\em the complex tensor product} of $n$ such 
algebras \Eq{oxM2C} 
is a standard tool for the description of the quantum circuits. 
On the other hand, {\em the real skew tensor product} of the same 
algebras \Eq{Cl3prod} is very natural for the construction of
Clifford algebra $\Cl_+(3n)$ and $Spin(3n)$ group.

However, in the usual tensor product all complex 
structures for different factors may be merged together,
but in the skew tensor product 
different complex units are anti-commuting and form 
a new term identified with $\Cl(n)$ in \Eq{ClHn}. 
The distinction between quantum circuits 
with $n$ qubits and $Spin(3n)$ group becomes 
natural with such a difference 
of the tensor products.

The description of $Spin(3n)$ by quantum circuits with
$2n$ qubits in \Sec{QCirc} provides an alternative approach. 
Here additional $n$ qubits are used for implementation of 
the anti-commuting structure. 

The quantum circuit model also may be used for 
a `physical' explanation of 
the complexity reduction for `$Spin(3n)$ circuits.'
New anti-commutative terms change 
commutation relations for Hamiltonians implementing partially
overlapped gates and (in agreement with Lie-algebraic 
approach to quantum circuits) prevent construction of 
gates of higher order from simpler gates, see \Sec{QCdec3n}.

Methods of classical
simulations of these quantum circuits with $2n$
qubits are discussed in the last section.
Due to such possibility $Spin(3n)$ group
is associated with {\em the new kind of quantum 
circuits which may be effectively simulated on 
a classical computer}.

Such a circuit uses only $n$ ({\em `primary'}) qubits
with even indexes as direct carriers of quantum information. 
Any gate with a single `primary' qubit may be realized, but  
an arbitrary entangled transformation for two `primary' qubits 
also involves a pair of ({\em `auxiliary'}) qubits 
with odd indexes and formally corresponds to a four-qubit gate, 
see \Eq{Hs6}.

For any pair of `primary' qubits the accompanying action 
may be dropped by the specific choice of a state for the 
`auxiliary' pair \Eq{UPsiUps}. However, for the 
composition of gates with overlapped `primary' qubits such 
side-effects may not be omitted, just because they ensure the 
significant reduction of the complexity.


\newpage
\appendix
\section{Computer algebra program}
\label{AppProg}

The {\tt OpenAxiom CAS} program\footnote{Latest version of {\tt Axiom CAS} 
may not work with the program due to a technical issue with a maximal number
of arguments in {\tt LISP} functions and {\tt OpenAxiom 1.4.1}
was used instead.}\: for conversion of a two-qubit gate into element 
of $Spin(6)$ group is presented below. 

Function {\tt MatClp6} calculates element of $S \in Spin(6)$ group
for $4\times 4$ unitary matrix of a two-qubit gate, after that 
action of $SO(6)$ group is calculated
using map $\mathbf{v} \mapsto S^{-1} \mathbf{v} S$.

\medskip

\begin{verbatim}
-- File SP6QC.input

K := Fraction Polynomial Integer
CK := Complex Fraction Integer
MK := Matrix Complex Fraction Integer
MK2 := SquareMatrix(2,CK)
MK4 := SquareMatrix(4,CK)

-- Pauli matrixes
sg0 : MK2 := matrix[[1,0],[0,1]]
sg1 : MK2 := matrix[[0,1],[1,0]]
sg2 : MK2 := matrix[[0,-%i],[%i,0]]
sg3 : MK2 := matrix[[1,0],[0,-1]]

-- Tensor products of Pauli matrixes
se00 : MK4 := tensorProduct(sg0,sg0)
se10 : MK4 := tensorProduct(sg1,sg0)
se20 : MK4 := tensorProduct(sg2,sg0)
se30 : MK4 := tensorProduct(sg3,sg0)
se01 : MK4 := tensorProduct(sg0,sg1)
se02 : MK4 := tensorProduct(sg0,sg2)
se03 : MK4 := tensorProduct(sg0,sg3)

-- Writing them into array
se:= vector([se10,se20,se30,se01,se02,se03])

-- Definition of Clifford algebra Cl_+(6)
Clp6 := CliffordAlgebra(6, K, quadraticForm diagonalMatrix[1,1,1,1,1,1])

-- Definition of generators
ep1 : Clp6 := e(1)
ep2 : Clp6 := e(2)
ep3 : Clp6 := e(3)
ep4 : Clp6 := e(4)
ep5 : Clp6 := e(5)
ep6 : Clp6 := e(6)

-- Writing them into array
ep := vector([ep1,ep2,ep3,ep4,ep5,ep6])

-- Definition of product of all generators: e1 e2 e3 e4 e5 e6
epi := ep1*ep2*ep3*ep4*ep5*ep6

-- Conversion of complex number: a + b i -> a + b e1 e2 e3 e4 e5 e6
CClp6 : (CK) -> Clp6
CClp6 z == real(z)+imag(z)*epi

-- Conversion of 4x4 complex matrices into real Clifford algebra Cl_+(6)
-- SU(4) matrix maps into element of Spin(6) group
MatClp6 : (MK4) -> Clp6
MatClp6 Mat ==
 Cl : Clp6 := CClp6(trace(Mat*se00)/4)
 for i in 2..6 repeat
  for j in 1..i-1 repeat
   if quo(i-1,3) = quo(j-1,3) then
       Cl := Cl + CClp6(trace(Mat*se(i)*se(j))/4)*ep(i)*ep(j)
      else
       Cl := Cl - CClp6(%i*trace(Mat*se(i)*se(j))/4)*ep(i)*ep(j)
 Cl


-- Unitary matrix
Uc := matrix[[1,0,0,0],[0,1,0,0],[0,0,0,-1],[0,0,1,0]]

-- Element of Spin(6) group
SPc := MatClp6(Uc)

-- Element of SO(6)
vec := a*ep1+b*ep2+c*ep3+d*ep4+e*ep5+f*ep6

-- Rotation of vec
recip(SPc) * vec * SPc

-- End of file SP6QC.input --
\end{verbatim}
 
\end{document}